\documentclass[pra,twocolumn,superscriptaddress,amssymb, floatfix,longbibliography]{revtex4-2}

\usepackage{graphicx}
\usepackage[usenames, dvipsnames]{color}
\usepackage{hyperref}
\hypersetup{
    colorlinks=true,
    linkcolor=Blue,
    citecolor=Blue,
    filecolor=Blue,      
    urlcolor=Blue,
    pdftitle={Neural-network quantum states for a two-leg Bose-Hubbard ladder under magnetic flux},
    }
\usepackage[all]{hypcap}
\usepackage{physics}

\begin{document}

\title{Neural-network quantum states for a two-leg Bose-Hubbard ladder under magnetic flux}

\author{K. \c{C}even}
\email{kadir.ceven@bilkent.edu.tr}
\affiliation{Department of Physics, Bilkent University, Ankara, 06800, Turkey}

\author{M. \"O. Oktel}
\affiliation{Department of Physics, Bilkent University, Ankara, 06800, Turkey}

\author{A. Kele\c{s}}
\affiliation{Department of Physics, Middle East Technical University, Ankara, 06800, Turkey}

\date{\today}

\begin{abstract}
Quantum gas systems are ideal analog quantum simulation platforms for tackling some of the most challenging problems in strongly correlated quantum matter. However, they also expose the urgent need for new theoretical frameworks. Simple models in one dimension, well studied with conventional methods, have received considerable recent attention as test cases for new approaches. Ladder models provide the logical next step, where established numerical methods are still reliable, but complications of higher dimensional effects like gauge fields can be introduced. In this paper, we investigate the application of the recently developed neural-network quantum states in the two-leg Bose-Hubbard ladder under strong synthetic magnetic fields. Based on the restricted Boltzmann machine and feedforward neural network, we show that variational neural networks can reliably predict the superfluid-Mott insulator phase diagram in the strong coupling limit comparable with the accuracy of the density-matrix renormalization group. In the weak coupling limit, neural networks also diagnose other many-body phenomena such as the vortex, chiral, and biased-ladder phases. Our work demonstrates that the two-leg Bose-Hubbard model with magnetic flux is an ideal test ground for future developments of neural-network quantum states.
\end{abstract}

\maketitle

\section{\label{sec:level1}Introduction}

Quantum simulation emerged as a powerful tool not only to realize long-sought practical technologies related to quantum information and quantum computation but also to study strongly correlated quantum matter \cite{gross2017quantum,daley2022practical}. Recent experimental progress established the quantum gas systems as the ideal analog quantum simulation platforms for tackling some of the most challenging problems in condensed matter from a fresh perspective which has the potential to elucidate the mysteries in superconducting cuprates, fractional quantum Hall systems, and frustrated quantum magnets \cite{Quintanilla_2009,RevModPhys.80.885,schafer2020tools, PhysRevA.81.053628}. Pioneering cold-atom experiments have already started to probe low-energy quantum correlations but have also revealed the urgent need for more reliable theoretical frameworks that can efficiently benchmark the experimental output \cite{hart2015observation}. 

A successful program that has received both experimental and theoretical prominence to alleviate this profound challenge is the investigation of toy models in reduced dimensional systems such as two-leg ladders. On the one hand, these are engineered experimental setups small enough to deploy the state-of-the-art accurate theoretical and numerical approaches developed for quasi-one dimensional systems; on the other hand, they are large enough to accommodate magnetic flux and the resulting complex many-body phases ---such as the vortex and chiral phases of the superfluid state akin to phases of superconductors under a magnetic field as well as the Mott insulating phase emanating from strong interactions \cite{Fisher_1989,greiner_2002}. Two-leg ladders are realized experimentally in a wide variety of cold-atomic systems 
\cite{Aidelsburger_2011, 
Aidelsburger_2013,  
Atala_2014, 
tai2017microscopy, 
stuhl2015visualizing,
Lin_2009,
Miyake_2013,
PhysRevLett.112.043001}, including synthetic dimensions \cite{PhysRevLett.112.043001, mancini2015observation,PhysRevA.94.063632,PhysRevA.105.043306}, and their theory is studied extensively in conjunction with critical experimental advances \cite{Orignac_2001,
Dhar_2012,Petrescu_2013,
Piraud_2015,
Tokuno_2014,
Keles_2015,
PhysRevA.94.063628,
PhysRevA.105.033303}.
In the ensuing effort to realize quantum simulation in a fully two-dimensional system and to build new theoretical schemes that can reliably predict the underlying physical phenomena, it is crucial to test the emerging numerical techniques in such well-controlled toy systems and expose their limits.

Variational and projection Monte Carlo techniques are among the most effective unbiased methods in the numerical studies of strongly correlated matter, especially in high dimensions. They have been widely utilized to study the Hubbard, $t\mathrm{-}J$, and Heisenberg models based on stochastic minimization of a class of variational wave functions derived from well-understood physical phenomena \cite{PhysRev.138.A442,PhysRevLett.80.4558,Becca_Sorella_2017,PhysRevX.5.041041}. In 2017, Carleo and Troyer proposed an alternative family of wave functions derived from neural networks trained in a similar variational Monte Carlo scheme, motivated by the remarkable progress of machine learning and artificial intelligence \cite{carleo_2017,RevModPhys.91.045002}. Their study laid the groundwork to demonstrate that these wave functions---dubbed the restricted Boltzmann machine ansatz or, more broadly, the neural-network quantum states---are capable of approximating the ground state and the dynamics of canonical strongly correlated quantum systems with polynomial resources in the exponential Hilbert space. Furthermore, their accuracy can be improved systematically, competing with some of the most sophisticated methods, such as tensor networks and projected entangled pair states (PEPS). Subsequently, it was shown that the neural-network quantum states contain volume law entanglement and have an expressive capacity analogous to the tensor network quantum states \cite{PhysRevX.7.021021,
gao2017efficient,
PhysRevB.99.155136,
PhysRevB.97.085104,
PhysRevB.97.035116,
PhysRevLett.122.065301,
2103.10293,
PhysRevLett.127.170601}.
Neural-network quantum states can be efficiently optimized using well-developed tools in machine learning and variational Monte Carlo techniques. These optimization methods alleviate the fundamental challenges in studying tensor networks, such as the difficulty of tensor contraction \cite{PhysRevLett.113.160503} or the exponential scaling of the matrix product state (MPS) bond dimension with transverse system size \cite{yan2011spin,PhysRevB.106.L081111,melko2019restricted}. Neural-network quantum states have been extensively generalized to different deep learning architectures and have been successfully applied to a wide range of problems in condensed matter physics 
\cite{PhysRevLett.121.167204,
PhysRevB.100.125124, 
PhysRevB.100.125131,
PhysRevResearch.2.033075,
PhysRevX.11.031034,
PhysRevX.11.041021}.

Motivated by these parallel developments in cold atomic systems and variational quantum Monte Carlo simulations, we use neural-network quantum states to probe the novel quantum phases that can be realized in two-leg Bose-Hubbard ladders under synthetic magnetic fields. Despite the significant interest from a fundamental theoretical point of view and the flurry of recent experimental progress, applications of neural networks to bosonic systems are relatively scarce. Before this work, they have been applied to study superfluid-Mott insulator transition in the Bose-Hubbard model using a restricted Boltzmann machine \cite{McBrian_2019} and a feedforward neural network ansatz \cite{Saito_2017,saito_2018}. Still, their efficiency was not investigated under artificial magnetic fields, which breaks time-reversal invariance and frustrates the many-body system. Our work aims to be a step toward filling this gap and to contribute to the explorations of alternative numerical schemes that can be useful in future studies of optical lattice experiments with synthetic magnetic fields. 
Furthermore, we also bring the two-leg flux ladder system and its surprisingly wide variety of many-body phases to the attention of neural network studies and showcase the potential of this system as a prototypical many-body system for future algorithmic developments. 

This paper is organized as follows: In Sec.~\ref{sec:NN}, we briefly review the restricted Boltzmann machine and feedforward neural network for variational Monte Carlo calculations and discuss some details about their application to the Bose-Hubbard model. In Sec.~\ref{sec:two_leg_ladder_BHM},
we introduce the two-leg Bose-Hubbard ladder under an artificial magnetic field and show how neural-network quantum states can reliably describe the canonical superfluid and insulator phases of this model.
First, in the strong coupling regime, we show that the neural-network quantum states can successfully capture the superfluid-Mott insulator transition in systems under magnetic flux in close agreement with the previous theoretical and numerical results \cite{Keles_2015}. Second, we obtain the chiral and vortex phases in the weak coupling limit, which are predicted theoretically  \cite{Orignac_2001} and confirmed experimentally \cite{Atala_2014}.
Lastly, in the latter regime, we confirm the existence of a novel quantum phase called the biased-ladder phase
using the neural networks initially predicted in \cite{Wei_2014}. 
In Sec.~\ref{sec:conclusions}, we give a summary and our main conclusions.

\section{\label{sec:NN}Neural-network quantum states}

In this section, we briefly review the neural network quantum states in the context of the Bose-Hubbard model and their stochastic optimization for completeness. We consider two different neural network architectures to calculate the ground state wave function of our model: The first is the restricted Boltzmann machine (RBM) ansatz, which was initially introduced in the seminal work of Carleo and Troyer \cite{carleo_2017} to describe quantum spin models, and also applied to study the phase diagram of one-dimensional Bose-Hubbard model \cite{McBrian_2019}. The second is the so-called feedforward neural network (FNN) which is well known in machine learning and has been recently applied to find the ground state of Bose-Hubbard model \cite{Saito_2017,saito_2018}.

\begin{figure}[t]
\includegraphics[width=\columnwidth]{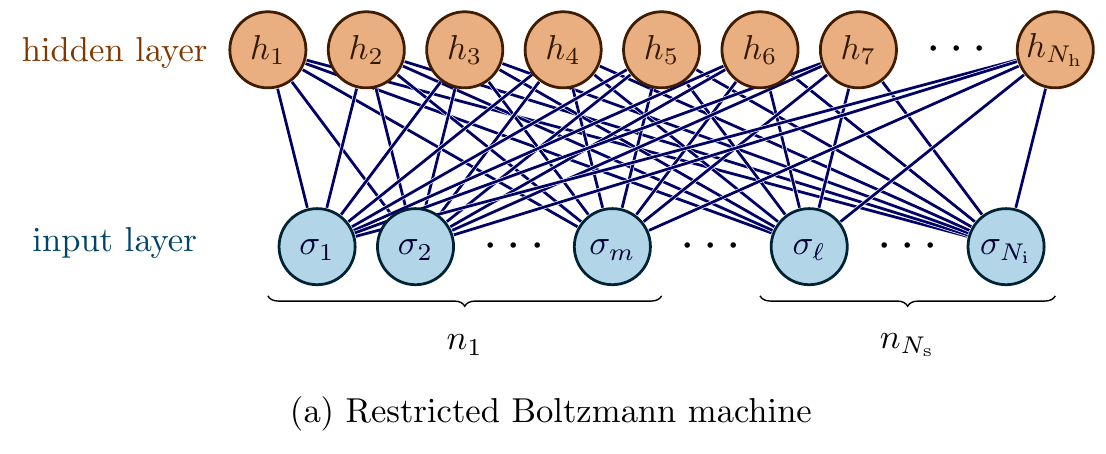}
\hfill
\includegraphics[width=\columnwidth]{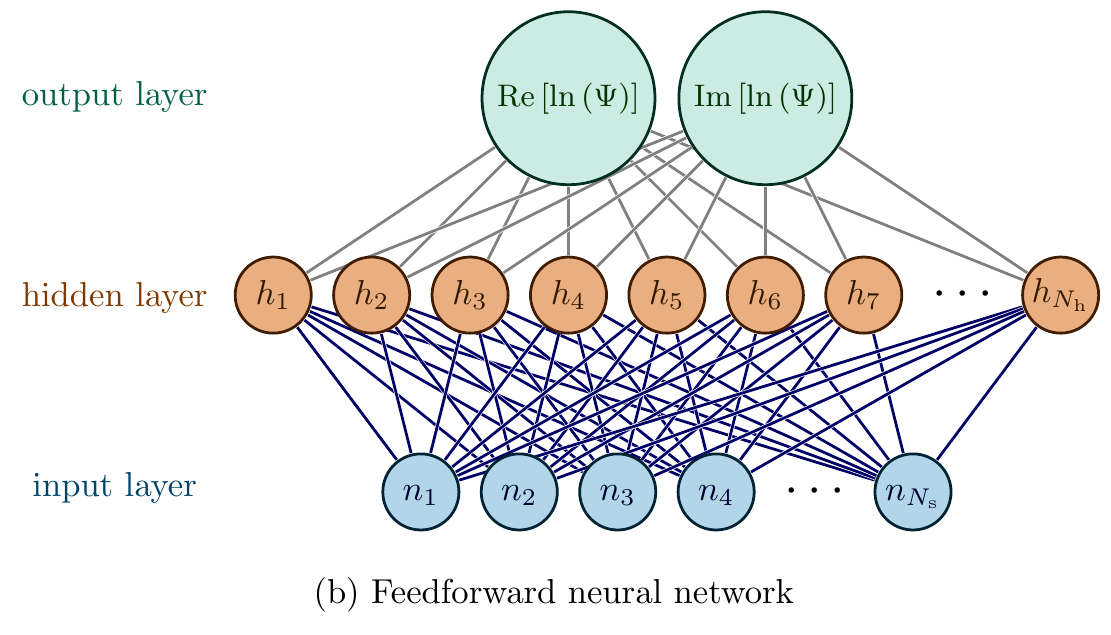}
\label{fig:RBMandFNN}
\caption{Restricted Boltzmann machine (a)  and feedforward neural network (b) applied to the Bose-Hubbard model. For (a), the input layer $\mathcal{S} = \left\{\,\sigma_j \mid \sigma_j \in \{\, 0, 1 \,\} \,\right\}_{j=1}^{N_\mathrm{i}}$ represents the physical space with the occupation number $n_k$, a so-called one-hot encoding with a maximum local occupation number $n_\mathrm{max}$. 
The number of input neurons is $N_\mathrm{i} = N_\mathrm{s} \times (n_\mathrm{max}+1)$. The hidden layer $\left\{\,h_i \mid h_i \in \{\, -1, 1 \,\} \,\right\}_{i=1}^{N_\mathrm{h}}$ contains $N_\mathrm{h}=\alpha \times N_\mathrm{i}$ neurons, where $\alpha \in \mathbb{Z}^+$. 
For (b), the input layer is composed of the site occupation numbers $\mathcal{S} = \{\, n_k \mid n_k \in \mathbb{Z}_0^+ \,\}_{k=1}^{N_\mathrm{s}}$, which have no limitation on the maximum occupation number. The hidden layer is composed of $N_\mathrm{h} = \alpha \times N_\mathrm{s}$ neurons where $\alpha \in \mathbb{Z}^+$. The output layer contains only two neurons, which are $\mathrm{Re}\left[ \ln (\Psi) \right]$ and $\mathrm{Im}\left[ \ln (\Psi) \right]$.}
\end{figure}

The RBM ansatz is constructed by considering input and hidden layers, just like the neural network studies in machine learning, as shown in Fig.~\hyperref[fig:RBMandFNN]{1(a)}. The input layer corresponds to the physical space. In contrast, the hidden layer is an abstract space determining the architecture of the variational parameters in the wave function. We use the so-called one-hot encoding for the bosonic occupation numbers in the physical space $\mathcal{S} = \left\{\,\sigma_j \mid \sigma_j \in \{\, 0, 1 \,\} \,\right\}_{j=1}^{N_\mathrm{i}}$ for a system of $N_\mathrm{i}$ sites in an arbitrary dimension. Each $k$th site in real space, where the total number of sites is $N_\mathrm{s}$, is represented with $n_\mathrm{max}+1$ binary elements $n_k=\{\,0, 0, \dots, 1, \dots, 0\,\}$ with only the $m$th element $1$ and all the others $0$, meaning that there are $m-1$ bosons on the $k$-th site. Here, we set an upper bound for the maximum occupation number per site as $n_\mathrm{max}$.
The hidden layer comprises $N_\mathrm{h}$ neurons, which take the values of $-1$ or $1$. Then, the variational wave function is written as a summation over the hidden layer neurons as follows
\begin{eqnarray}\label{eq:rbm_wf}
    \Psi(\mathcal{S};\mathcal{W}) =&& \sum_{\{h_i\}} e^{\sum_{i} a_i h_i + \sum_{j} b_j \sigma_j + \sum_{ij} W_{ij} h_i \sigma_j },
    \label{eq:rbm_visible_and_hidden}
\end{eqnarray}
which is the probability amplitude of the given state $\mathcal{S}$ and the dependence on the set of all variational parameters is shown with $\mathcal{W}=\{\,a_i,b_j,W_{ij}\,\}$ for $i=1,\dots,N_\mathrm{h}$ and $j=1,\dots,N_\mathrm{i}$. 
Conventionally, $a_i$ and $b_j$ are called bias parameters for input and hidden layers, respectively, and $W_{ij}$ are called the weights of the links between the layers.
Crucially, any neuron in the visible layer is connected to all neurons in the hidden layer. However, there is no connection between two neurons in the same layer, hence the name ``restricted.''
Thanks to this assumption, we can trace out the hidden neurons by performing the summation over $\{\,h_i\,\}$ in Eq.~\eqref{eq:rbm_visible_and_hidden}, which results in the following compact expression for the variational ansatz
\begin{eqnarray}\label{eq:rbm_wf_reduced}
    \Psi(\mathcal{S};\mathcal{W}) =&& e^{\sum_j b_j \sigma_j} \prod_i \theta_i (\mathcal{S};\mathcal{W}),
\end{eqnarray}
where
\begin{equation}\label{eq:theta_i}
    \theta_i (\mathcal{S};\mathcal{W}) = 2 \cosh{\left( a_i + \sum_j W_{ij} \sigma_j \right)}~.
\end{equation}
For $N_\mathrm{s}$ lattice sites, $N_\mathrm{h}$ hidden neurons, and a given maximum occupation number $n_\mathrm{max}$, we have $N_\mathrm{i} = (n_\mathrm{max}+1) \times N_\mathrm{s}$ 
input neurons.
We denote the hidden layer neuron density with $\alpha = N_\mathrm{h}/N_\mathrm{i} \in \mathbb{Z}^{+}$, which can be tuned to increase the accuracy of the ansatz.
Notably, the network parameters can be $\mathbb{R}$ and $\mathbb{C}$ valued in our implementation, which is essential for systems with broken time-reversal invariance.


In contrast, FNN ansatz first applied to Bose-Hubbard model in \cite{Saito_2017}, is composed of three layers, as shown in Fig.~\hyperref[fig:RBMandFNN]{1(b)}. The input layer corresponds to the physical sites as in RBM, but each neuron takes integer values corresponding to the number of bosons in that site without any cutoff in a maximum occupation such that $\mathcal{S} = \{\, n_k \mid n_k \in \mathbb{Z}_0^+ \,\}_{k=1}^{N_\mathrm{s}}$. 
The hidden layer, which consists of $\mathbb{R}$-valued $N_\mathrm{h}$ neurons, is obtained by
\begin{equation}
    h_j(\mathcal{S};\mathcal{V}) = b_j + \sum_{k} V_{j k} n_k,
\end{equation}
where
$\mathcal{V} = \{\, b_j, V_{j k} \,\}$ is the set of variational network parameters connecting the input and hidden layers, $b_j$ are the biases for the hidden layer, and $V_{j k}$ are the weights of links between the layers.
Lastly, the output layer contains only two neurons obtained from the hidden layer using the so-called hyperbolic tangent activation function, which amounts to
\begin{equation}
    u_i(\mathcal{S};\mathcal{W}) = a_i + \sum_{j} W_{i j} \tanh{ \left[ h_j(\mathcal{S};\mathcal{V}) \right] },
\end{equation}
where $i=1,2$. Here $\mathcal{W} = \{\, a_i, W_{i j}, b_j, V_{jk} \,\}$ is the combined set of all variational network parameters between the input and the output layers.
The final variational wave function is written in terms of the neurons in the output layer as 
\begin{equation}\label{eq:ffnn}
    \Psi(\mathcal{S}; \mathcal{W}) = e^{u_1(\mathcal{S}; \mathcal{W}) + i u_2(\mathcal{S}; \mathcal{W})},
\end{equation}
which is $\mathbb{C}$ valued as required for our system, whereas all the network parameters are $\mathbb{R}$ valued. This gives us a significant computational advantage compared to the RBM.
We use a similar definition for the hidden layer neuron density as $\alpha = N_\mathrm{h}/N_\mathrm{s}$.
In our implementation, these variational wave functions are conveniently created using \textsc{flax} \cite{flax_2020}, a neural network library for \textsc{jax} \cite{jax_2018}.

For sampling from the Hilbert space and optimizing the variational network parameters, 
we follow the same approach used in Ref.~\cite{carleo_2017}, which uses a Markov chain Monte Carlo (MCMC) based on the Metropolis-Hastings algorithm \cite{hastings_1970}. 
We start with randomly initialized variational parameters $\mathcal{W}$.
From a given configuration $\mathcal{S}^{(i)}\equiv\mathcal{S}$, we suggest a 
candidate configuration $\mathcal{S}^\prime$ and 
calculate the probability amplitudes from the neural network ansatz. 
If the acceptance ratio defined by
$p_\mathrm{ratio} = \abs{\Psi \left( \mathcal{S}^\prime; \mathcal{W} \right) / \Psi \left( \mathcal{S}; \mathcal{W} \right)}^2$ is greater than a random number sampled from the uniform distribution,
accept $\mathcal{S}^\prime$ as the next configuration $\mathcal{S}^{(i+1)}=\mathcal{S}^\prime$; otherwise, set $\mathcal{S}^{(i+1)}=\mathcal{S}$ in the Markov chain.
We implemented 
update rules both in canonical and grand canonical ensembles with and without conservation of the total number of particles, respectively, and confirmed the consistency of the converged ground states.
After many samples are generated in the Markov chain, we calculate the expectation values of the necessary observables, such as the ground state energy stochastically \cite{PhysRev.138.A442}, as follows:
\begin{eqnarray}\label{eq:average}
    \expval{A}=
     \frac{1}{M}\sum_{i=1}^M 
    A_\mathrm{loc}\left( \mathcal{S}^{(i)};\mathcal{W} \right),
\end{eqnarray}
where $M$ is the number of samples in the Markov chain and
$A_\mathrm{loc}$ is the so-called local observable defined by
\begin{equation}
    A_\mathrm{loc}\left( \mathcal{S}^{(i)} ;\mathcal{W}\right) = \sum_k \mel{i}{A}{k}  \frac{\Psi\left( \mathcal{S}^{(k)};\mathcal{W} \right)}{\Psi\left( \mathcal{S}^{(i)};\mathcal{W} \right)},
\end{equation}
which can be efficiently calculated in the entire Hilbert space for a given configuration $\mathcal{S}^{(i)}$ due to the sparsity of the most often required observables of interest.
More importantly, gradients of the energy, which will require the calculation of the observables set up from the derivatives of the ansatz in Eqs.~\eqref{eq:rbm_wf} and \eqref{eq:ffnn}, can also be calculated from the same Markov chain efficiently, enabling the use of gradient-based optimization with respect to variational parameters $\mathcal{W}$. 

Typically, even the most straightforward optimization schemes, like the method of steepest descent, can be used for elementary states, as demonstrated in \cite{Saito_2017}. For more complicated ground states requiring a larger set of variational parameters with slower convergence in a rugged energy landscape, one needs to implement more robust optimizations such as the stochastic reconfiguration method \cite{PhysRevLett.80.4558, Sorella_2001, Becca_Sorella_2017}, which was also applied to neural networks \cite{carleo_2017}.
Here, we use the adaptive moment estimation (\textsc{adam}) \cite{adam} as the optimizer instead of the stochastic reconfiguration to avoid the computationally expensive calculation of the inverse of the correlation matrix.
\textsc{adam} is a gradient-based optimizer that adaptively estimates the first and second moments of the energy derivatives for the optimization of neural network ansatz, 
which can be conveniently implemented using the gradient processing and optimization library \textsc{optax} \cite{optax_2020} that can be used on the top of powerful \textsc{jax} \cite{jax_2018}. We benchmarked our code in the one-dimensional Bose-Hubbard model and found the superfluid-Mott insulator phase diagram to be in excellent agreement with the density-matrix renormalization group (DMRG) result. Similarly, we tested our code in a two-leg Bose-Hubbard ladder with $L=32$, $\alpha=4$, $M=10^4$, and an \textsc{adam} learning rate of $0.01$ and determined that a phase convergence, especially for a biased-ladder phase, can take up to 28 hours with an Apple M1 CPU.

\section{\label{sec:two_leg_ladder_BHM}Two-leg Bose-Hubbard ladder in artificial magnetic field}
\begin{figure}[t]
\includegraphics[width=0.8\columnwidth]{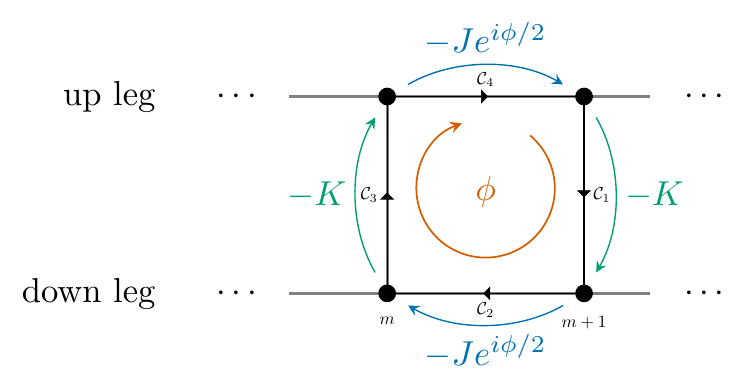}
\caption{\label{fig:tll_lattice} Definition of hopping amplitudes in two-leg Bose-Hubbard model under magnetic field.  Due to the applied artificial magnetic field, hopping amplitudes have Peierls phases, and each plaquette is pierced with flux $\phi$. }
\end{figure}

We now consider a bosonic system confined on an optical lattice in the form of ladder geometry with rungs, as shown in Fig.~\ref{fig:tll_lattice}.
The Hamiltonian is given by
\begin{eqnarray}\label{eq:tll_bhm}
    \mathcal{H} =&& 
    -J \sum_{m=1}^{L} \sum_{\ell \in \{\mathrm{u}, \dd\}} \left(e^{i \sigma_\ell\phi/2} a_{\ell, m+1}^\dagger a_{\ell, m} + \mathrm{h.c.}\right)\\
    && - K \sum_{m=1}^{L} \left(a_{\mathrm{u}, m}^\dagger a_{\dd, m} + \mathrm{h.c.}\right)\nonumber \\
    && + \sum_{m=1}^{L} \sum_{\ell \in \{\mathrm{u}, \dd\}} \left[\frac{U}{2} n_{\ell,m}\left(n_{\ell,m} -1\right) - \mu n_{\ell,m}\right]\nonumber,
\end{eqnarray}
where $a_{\ell,m}^\dagger$ and $a_{\ell,m}$ are bosonic creation and annihilation operators at the site $(\ell,m)$,
$\ell=\mathrm{u},\mathrm{d}$ labels upper and lower site in a rung, 
$n_{\ell,m}=a_{\ell,m}^\dagger a_{\ell,m}$ is the boson number operator,
$J$ is the intraleg hopping amplitude, $K$ is the interleg hopping amplitude,
$U$ is onsite interaction strength, $\mu$ is the chemical potential, 
$L$ is the number of rungs, 
$\phi$ is a phase from the artificial magnetic field and $\sigma_\ell=+1$ ($-1$) for $\ell=\mathrm{u}$ ($\mathrm{d}$). 
The hopping amplitudes $J$, $K$ and the overall phase $\phi$ can be separately engineered in a standard cold atomic system through adjusting the depth of the optical lattice and a more complicated Raman coupling or shaking technique \cite{Struck_2012,Aidelsburger_2013,Atala_2014,PhysRevLett.112.043001}. By using the Peierls substitution \cite{Hofstadter_1976}, one can relate the phase to an artificial magnetic flux passing through each plaquette
$\phi=\int_\mathcal{A} \vb{B} \cdot \dd\vb{a}= \oint_\mathcal{C} \vb{A} \cdot \dd \vb{l}$, where $\vb{B}$ and $\vb{A}$ are the corresponding synthetic magnetic field and vector potentials, respectively. $\mathcal{C}$ is a closed path around a plaquette; $\mathcal{A}$ is the area, and
$\vb{B} = \curl{\vb{A}}$. Unlike the standard solid state experiments, the phase here can take any value giving rise to arbitrarily large artificial magnetic flux in this system, which paves the way for many desired strongly correlated states.
Due to the invariance of the Hamiltonian in Eq.~\eqref{eq:tll_bhm} under the transformation $(\mathrm{u},\mathrm{d}, \phi) \rightarrow (\mathrm{d}, \mathrm{u}, -\phi)$, one can focus on the domain $0<\phi\leq\pi$. 
Above a critical field $\phi_\mathrm{c}$, the single-particle spectrum shows two degenerate minima at some $\pm k_\mathrm{c} \neq 0$, which is the key to obtaining novel many-body phases by tuning the interaction in this system. Here, we are interested in testing which phases can be captured reliably within the variational neural networks and in exposing their accuracy.
\begin{figure}[t]
\includegraphics[width=\columnwidth]{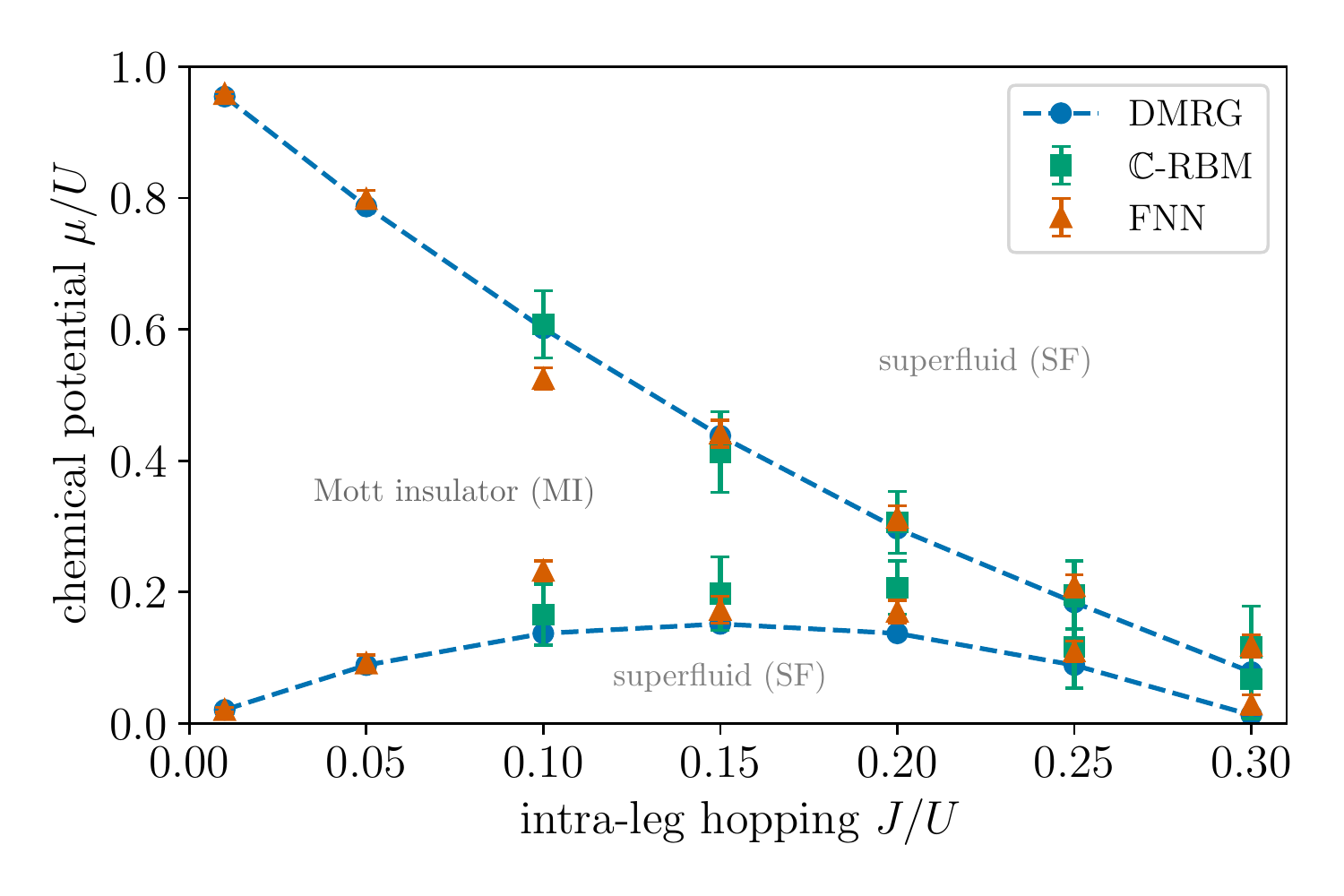}
\caption{\label{fig:tll_bhm_strong_phase_diag} Superfluid-Mott insulator phase diagram of the two-leg ladder Bose-Hubbard model with $K/J=1.00$ and $\phi/\pi = 0.90$ from the RBM (green square) and FNN (orange up-triangle) ansatzes for a system of $L=12$ sites  compared with DMRG (blue circle) \cite{Keles_2015}.}
\end{figure}

We first focus on the strongly interacting regime $J/U \ll 1$ to determine the phase boundaries of the Mott insulator state in the presence of finite flux as an initial benchmark of our implementation. 
As emphasized in the previous section, $\mathbb{C}$-valued RBM is necessary here since the Hamiltonian in Eq.~\eqref{eq:tll_bhm} breaks the time-reversal symmetry. 
In the FNN, the variational network parameters are $\mathbb{R}$ valued even though the final wave function in the output layer in Fig.~\hyperref[fig:RBMandFNN]{1(b)} has both real and imaginary parts. For a total number of sites $N_\mathrm{s}$, one can calculate the variational energies for $N_\mathrm{s}$ and $N_\mathrm{s} \pm 1$ bosons by optimizing the RBM and FNN ansatzes in the given filling sectors. Here, the energies with $N_\mathrm{s}\pm 1$ bosons correspond to the particle and hole excitation energies of the system.  
Then, one can write the following expression for the chemical potential, which is also known as the charge gap:
\begin{equation}\label{eq:particle–hole excitation_energies}
    \pm \mu_\pm = E(N_\mathrm{s} \pm 1) - E(N_\mathrm{s}),
\end{equation}
where the energies of $N_\mathrm{b}$ bosons on a fixed system of $N_\mathrm{s}$ lattice sites are represented with $E(N_\mathrm{b})$, leaving lattice size implicit for brevity. Each energy in Eq.~\eqref{eq:particle–hole excitation_energies} is calculated stochastically from the expression in Eq.~\eqref{eq:average} as the minimum energy in the given sector. The resulting phase diagram in the $\mu\mathrm{-}J$ plane is shown in Fig.~\ref{fig:tll_bhm_strong_phase_diag}, along with the DMRG results \cite{Keles_2015}. For small $J/U$, where a mean-field Gutzwiller ansatz is more reliable, RBM and FNN show excellent agreement with DMRG with minimal statistical fluctuations. For larger $J/U$, the deviations from the DMRG data and the enhanced statistical fluctuations are more pronounced. However, the results are still superior to any mean-field approach, showcasing neural-networks' capability of capturing basic correlations. The proximity of the Mott insulator phase is expected to facilitate more complicated correlated phases deep inside the superfluid phase due to competing effects of the magnetic field, kinetic energy, and strong interactions \cite{PhysRevA.76.055601} such as fractional quantum Hall phases \cite{PhysRevA.81.053628} and charge-density waves \cite{PhysRevA.90.053623}. Therefore, this region can be a good candidate for designing more innovative neural networks in future studies, which may also have direct experimental relevance in new generation cold-atom setups. Note that, around the tip of the Mott insulator phase, the transition is of Berezinskii-Kosterlitz-Thouless type driven by phase fluctuations rather than number fluctuations. It requires an analysis different from Eq.~\eqref{eq:particle–hole excitation_energies}, which is beyond the scope of this work \cite{Kuhner_1998}. 
\begin{figure*}[t]
\includegraphics[width=\columnwidth]{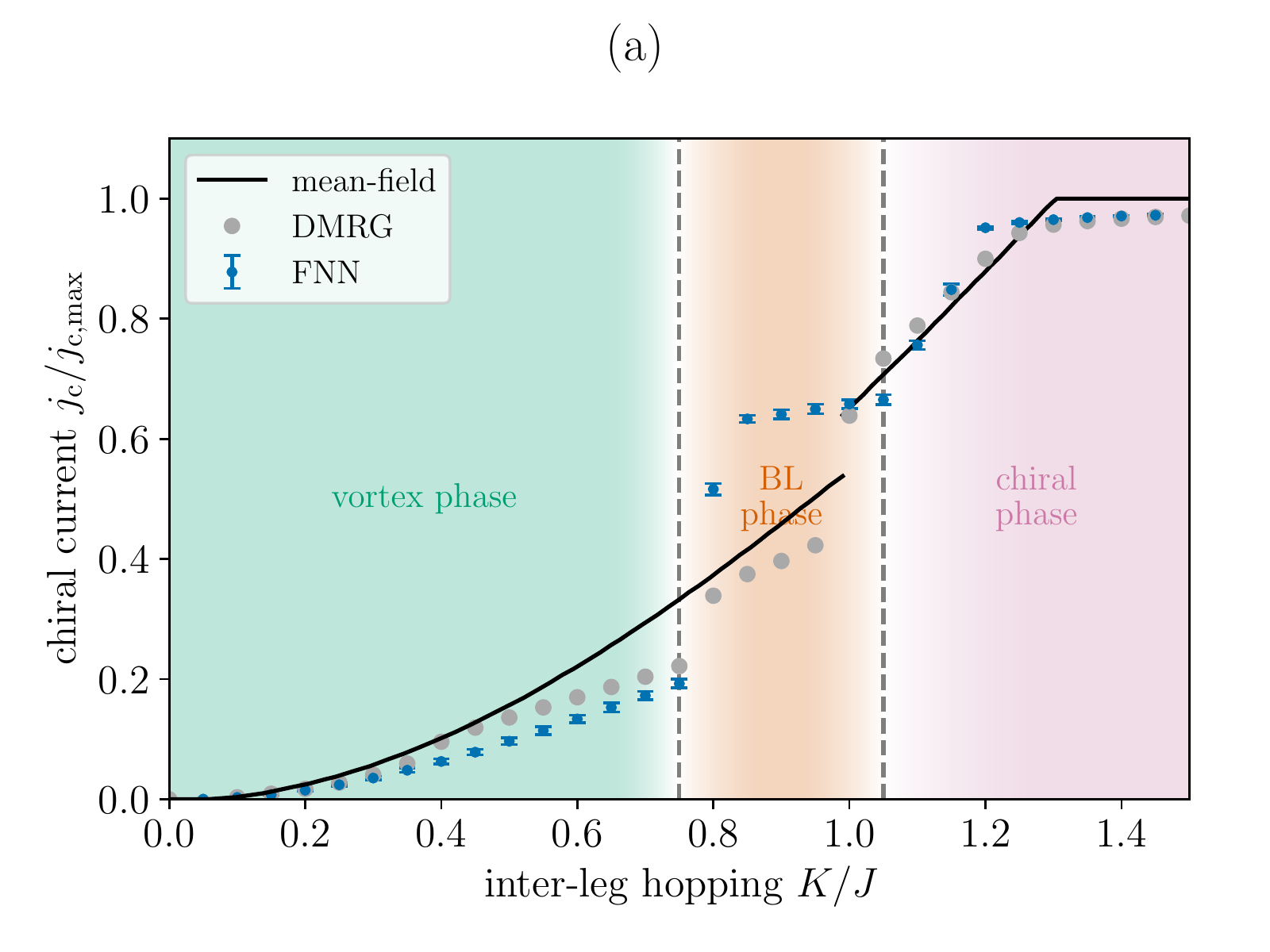}
\hfill
\includegraphics[width=\columnwidth]{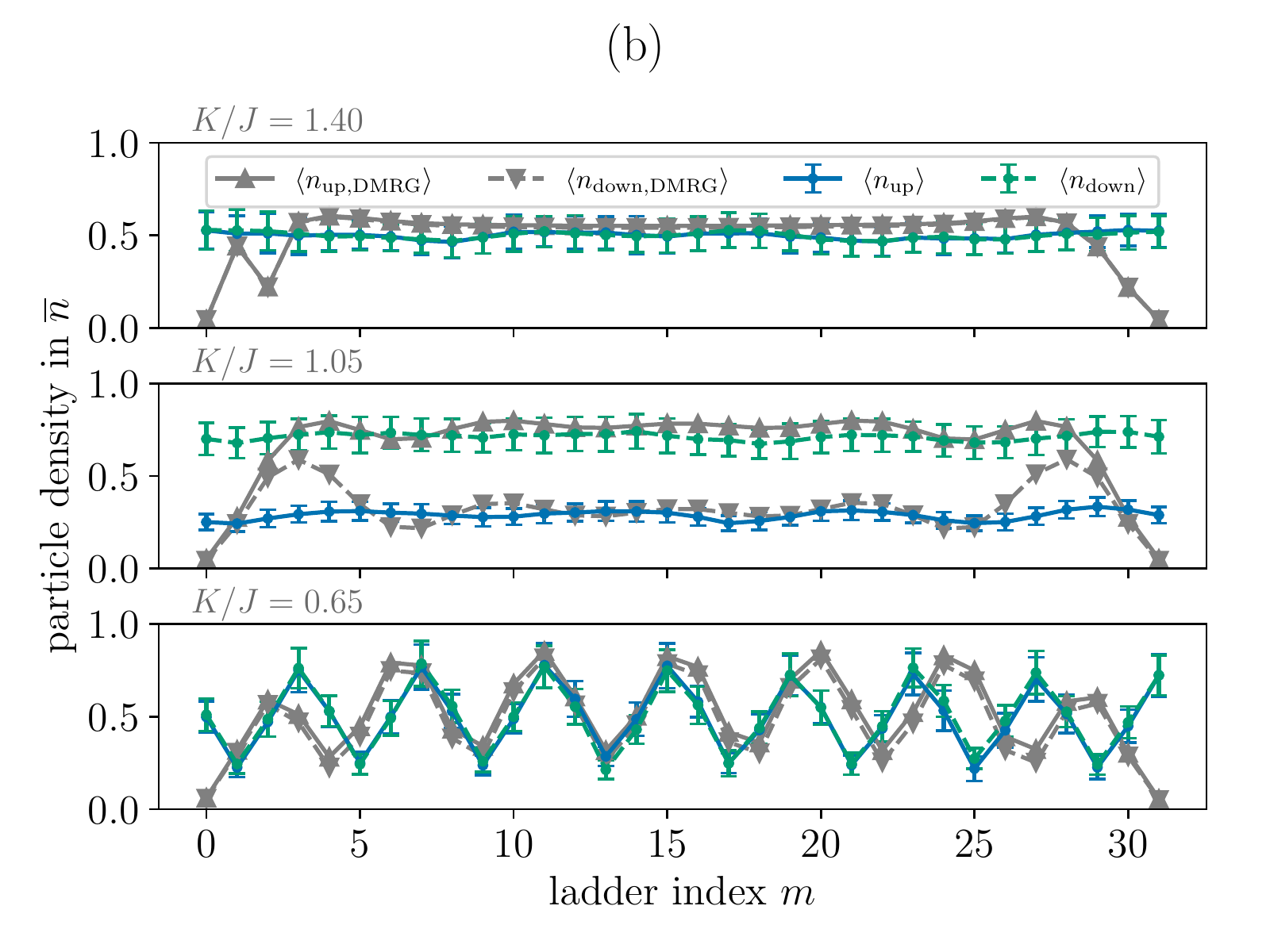}
\label{fig:tllFNN}
 \caption{(a) The chiral current as a function of $K/J$ for fixed $N=L=16$, $U/J=0.20$, and $\phi/\pi=0.50$ from the FNN ansatz (blue dot) normalized by the maximum current $j_{\mathrm{c},\mathrm{max}}=2\overline{n}J \sin(\phi/2)=J\sqrt{2}$ compared with the mean-field result for $L=32$ (black line) \cite{Wei_2014} and DMRG for $L=64$ (dark-gray circle).
The phase diagram has vortex, biased-ladder, and chiral phases. The dashed lines show the approximate phase boundaries. (b) The particle density normalized by the mean density $\overline{n} = N/L$ for $L=32$ as a function site index $m$ in upper (blue dot with line) and lower (green dot with dashed line) legs for selected points in the phase diagram $K/J=1.40$ (top), $K/J=1.05$ (middle), and $K/J=0.65$ (bottom) compared with DMRG (up- and down-triangles).}
\end{figure*}

We now move on to the regime of weaker interactions $J/U \gg 1$ outside the Mott insulator to probe the states with superfluidity based on the FNN ansatz exclusively, which we found to be more accurate. A surprisingly wide variety of superfluid phases can be realized in this system with different vortex and particle density profiles \cite{PhysRevA.94.063628}. For simplicity, we focus on three basic phases that can be tuned as a function of interleg hopping $K$ for a fixed magnetic flux.
In the regime $K/J\ll 1$, where the legs are coupled weakly, one has the vortex phase on which the particle density along the ladders is modulated, its period increases with $K/J$, and there are currents along the rungs. 
In regime $K/J\approx 1$, where the legs are coupled strongly, one has the chiral phase, also called the Meissner phase. Here, the particle density is homogeneous along and across the legs. The superfluid velocities are equal and opposite across the legs, with no flow from one leg to the other along the rungs.
In the intermediate regime, one has the biased-ladder (BL) phase, where  the particle density is larger in one leg and the superfluid velocity is larger in the other leg, but the net current still is zero.
We identify these phases by calculating the densities at each site and the currents along and across the legs.
The current operators along the legs and rungs are given as 
\begin{eqnarray}
    j^\parallel_{\ell, m} =&& i J\left( e^{i\sigma_\ell\phi/2} a_{\ell, m+1}^\dagger a_{\ell, m} - \mathrm{h.c.} \right)\\
    j^\perp_m =&& iK\left( a_{\mathrm{u}, m}^\dagger a_{\dd, m} - \mathrm{h.c.}\right)~.
\end{eqnarray}
Here $j^\parallel_{\mathrm{u},m}$ and $j^\parallel_{\mathrm{d},m}$ are the currents along the upper and lower legs, respectively. 
The chiral current operator is defined as
\begin{eqnarray}
    j_\mathrm{c} = \frac{1}{L}\sum_m \left( j^\parallel_{\mathrm{u}, m} - j^\parallel_{\mathrm{d}, m}\right)~.
\end{eqnarray}
The maximum value of the chiral current is $j_{\mathrm{c},\mathrm{max}}=2\overline{n}J \sin(\phi/2)$ \cite{Wei_2014}, where $\overline{n}=N/L$ is the mean density and $N$ is the number of particles. This will be used for the overall normalization below such that in the chiral phase $\expval{j_\mathrm{c}}\approx j_{\mathrm{c},\mathrm{max}}$, and in other phases $\expval{j_\mathrm{c}}<j_{\mathrm{c},\mathrm{max}}$.

\begin{figure*}[t]
\includegraphics[width=\columnwidth]{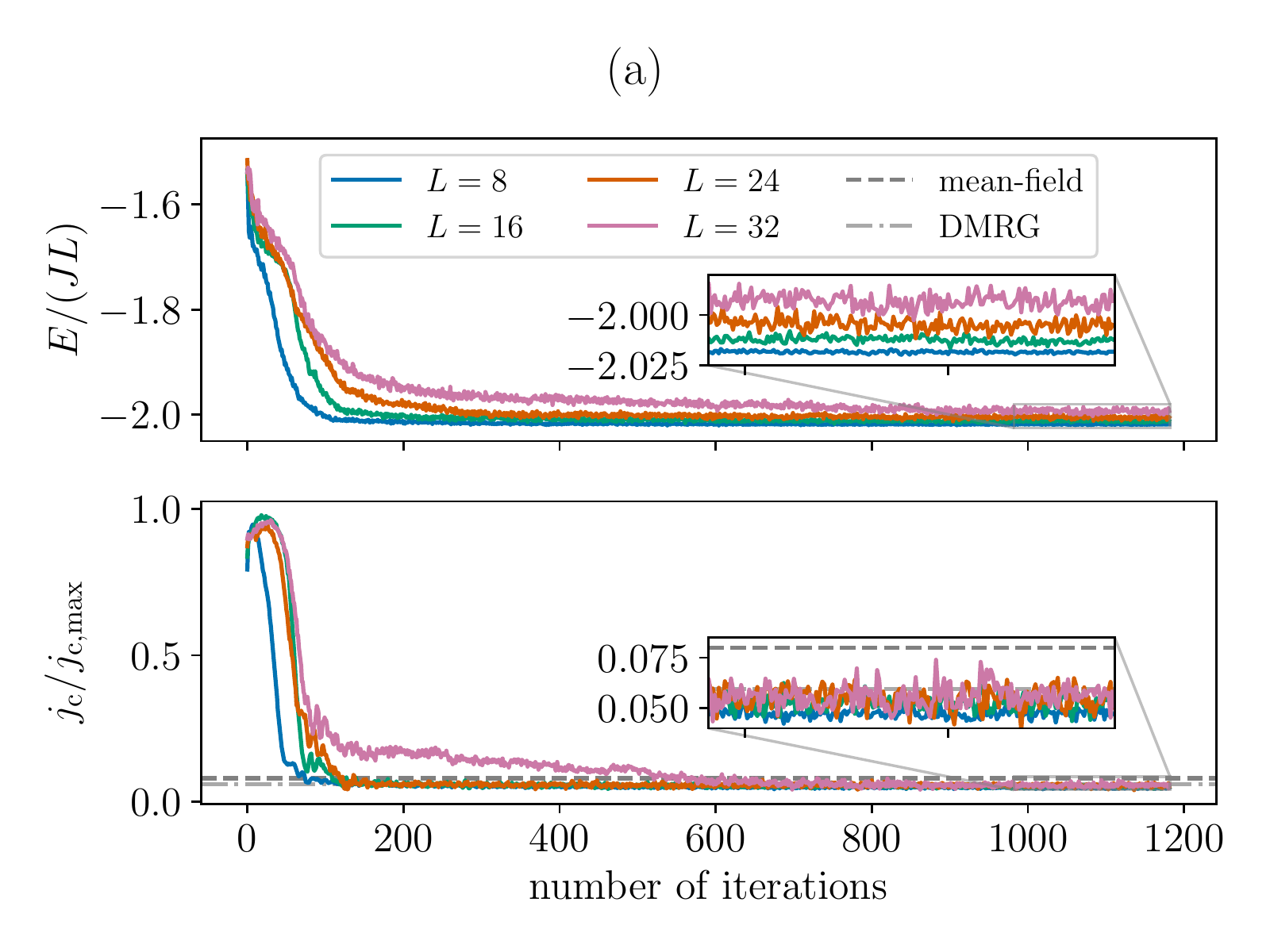}
\hfill
\includegraphics[width=\columnwidth]{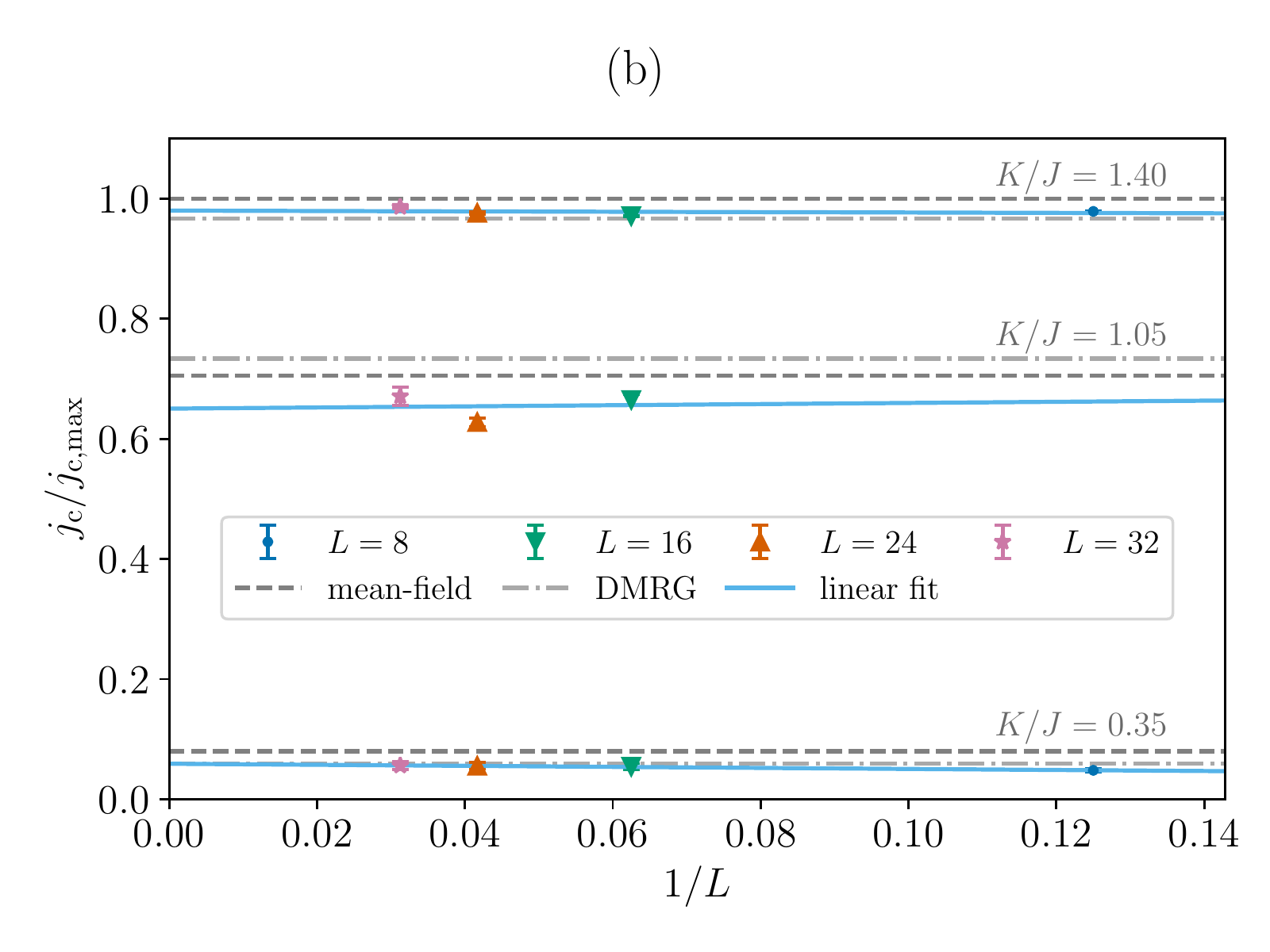}
\label{fig:energy_current_vs_iter_and_1_over_L}
 \caption{(a) The variational energy normalized by the intraleg hopping amplitude $J$ times the number of rungs $L$ and the normalized chiral current as a function of the number of \textsc{adam} iterations for fixed $K/J=0.35$, $U/J=0.20$, and $\phi/\pi=0.50$ from the FNN ansatz in $L=8$ (blue [dimgray] line), $16$ (green [gray] line), $24$ (orange [dark gray] line), $32$ (pink [silver] line) compared with the mean-field result for $L=32$ (dashed line) \cite{Wei_2014} and DMRG for $L=64$ (dash-dot line). (b) The normalized chiral current as a function of the inverse number of rungs $1/L$ compared with the mean-field result \cite{Wei_2014} and DMRG. The light blue lines indicate the linear fitting of chiral currents over $1/L=8^{-1}$ (blue dot), $16^{-1}$ (green down-triangle), $24^{-1}$ (orange up-triangle), $32^{-1}$ (pink star) for the selected points in the phase diagram $K/J=1.40$ (top), $K/J=1.05$ (middle), and $K/J=0.35$ (bottom).}
\end{figure*}

In Fig.~\hyperref[fig:tllFNN]{4(a)}, we present a cut along the phase diagram as a function of $K/J$ for fixed $U/J=0.20$, and $\phi/\pi=0.50$ from the expectation value of the chiral current for $N=L=16$ and particle densities for $N=L=32$. In the vortex phase (green region), the chiral current is small and grows smoothly to a finite value much smaller than $j_\mathrm{c,max}$, whereas the particle densities, shown in Fig.~\hyperref[fig:tllFNN]{4(b)}, are equal in the upper and lower legs and oscillate along the legs. Above a critical $K/J\approx 0.75$, the chiral current shows a rapid increase to a larger value which is still less than $j_\mathrm{c,max}$. Within this intermediate BL phase (orange region), the current continues to grow slowly, and the particle densities at the upper and lower legs are different. After a second critical point $K/J\approx 1.05$, inside the expected chiral phase (purple region), the chiral current saturates close to $j_{\mathrm{c}, \mathrm{max}}$ and the particle densities, which are equal in upper and lower sites, become uniform along the legs.
It is important to note that convergence of the energy is considerably slow within the BL phase due to delicate competition between the vortex and chiral phases. Whereas the regions around the Mott insulator phases, as well as the vortex and chiral phases, took at most a few thousand iterations for convergence (see Fig.~\ref{fig:energy_current_vs_iter_and_1_over_L}), we observed a slow drift of energy in the BL phase up until 60-70 thousand iterations and the density profiles changed considerably before converging to the BL phase. The RBM ansatz, which is not shown here, also performed relatively poorly around this region. Finding more competitive network architectures to capture this region more accurately is the subject of future work. Nevertheless, the fact that the FNN ansatz could reveal this phase without any bias still demonstrates the power of neural-network quantum states in finding novel quantum many-body phases.

For benchmarking, we have also included the chiral current calculated using the state-of-the-art DMRG simulation for $L=64$ \cite{itensor}. For these, we took MPS bond dimensions up to 160 and performed 15 sweeps for each data point. To reduce the effect of initial random MPS and prevent DMRG from being trapped in local minima, we increased $K$ from zero gradually and used the optimized MPS obtained for a smaller $K$ for the initial ansatz of the next value $K+\delta K$. One can see from Fig.~\hyperref[fig:tllFNN]{4(a)} that FNN correlates with DMRG in the vortex phase region, whereas the mean-field results give a much higher chiral current than these two simulations. In the chiral phase region, nearly all the results agree with each other. However, in the biased-ladder phase region, the currents start to vary from the DMRG more markedly, which is expected since DMRG uses open boundary conditions, which greatly affect the chiral current due to boundaries.

\section{\label{sec:conclusions}Summary and Conclusions}

In this paper, we developed an application of neural-network quantum states to find the many-body phases in the two-leg ladder Bose-Hubbard model under an artificial magnetic field. The strong magnetic flux, inter- and intraleg kinetic hopping, and the onsite interactions enable a plethora of competing phases in this toy model, which is ideal for testing the power of neural networks. Due to the broken time-reversal invariance, we implemented RBM ansatz with complex network parameters and the FNN with real parameters but two output neurons for real and imaginary parts of wave function amplitude. In the strong coupling regime where onsite interactions dominate, we showed that both RBM and FNN wave functions describe the Mott transition reliably, in a precision comparable with the DMRG results. In the weak coupling regime with competing superfluid phases, we focused on the FNN wave function. We showed that three phases are predicted from the FNN ansatz. First is the vortex phase in the weak leg coupling regime, where particle density has modulations and homogeneous superflow. Second, the chiral phase with uniform density and leg currents in opposite directions. Third, the biased-ladder phase where particle density and superfluid velocities are different in each leg, but the total current is zero. We emphasize that these phases came out of the variational minimization of the ground-state energies without bias.

Our work demonstrates that the two-leg ladder Bose-Hubbard model with magnetic flux is an ideal test ground for future developments of neural-network quantum states. Several key questions can be investigated with other promising neural networks. First, the vicinity of the Mott insulator phase where the phase boundary showed enhanced fluctuations is a candidate for strongly correlated phases \cite{Keles_2015,PhysRevA.94.063628}, which may be unveiled by the use of more sophisticated networks such as the recently developed convolutional networks, which show improved accuracy \cite{PhysRevB.100.125124,roth2021group}. This regime is also suitable for experimental investigations in the new generation quantum gas setups. Secondly, even for weak interactions where different superfluid phases compete, the convergence of the neural network showed considerable slowing down, which can be circumvented in alternative ansatzes. However, the FNN can reasonably give the expected biased-ladder phase. Apart from the realization of this system in cold atom experiments with a high degree of control, the ability to benchmark this toy model with independent powerful numerical tools like DMRG or quantum Monte Carlo using worm sampling \cite{PhysRevA.77.015602} makes future studies of this system beneficial for both techniques.

\begin{acknowledgments}
A.K. is supported by  T{\"U}B{\.I}TAK 2236 Co-funded Brain Circulation Scheme 2 (CoCirculation2) Project No. 120C066.
\end{acknowledgments}

\bibliography{references.bib}

\end{document}